\documentclass[prl,reprint,superscriptaddress,twocolumn]{revtex4-1}
\usepackage{graphicx}
\usepackage{blindtext}
\usepackage{xcolor}
\usepackage{lineno}
\usepackage{amsmath}
\usepackage{todonotes}
\usepackage{comment}
\definecolor{revblue}{cmyk}{0.95,0.85,0.01,0.04}
\usepackage[colorlinks=true,
            linkcolor=black,
            urlcolor=revblue,
            citecolor=revblue]{hyperref}

\newcounter{mnotecount}[section]
\renewcommand{\themnotecount}{\thesection.\arabic{mnotecount}}
\newcommand{\mnotex}[1]%{}
{\protect{\stepcounter{mnotecount}}$^{\mbox{\footnotesize
$%\!\!\!\!\!\!\,
\bullet$\themnotecount}}$ \marginpar{%\color{red}%
\raggedright\tiny\em 
$\!\!\!\!\!\!\,\bullet$\themnotecount: #1} }

\usepackage{chngcntr}
\counterwithout{equation}{section}

\begin{document}
%\lipsum
\title{Levitodynamic spectroscopy for single nanoparticle characterization}
\author{Jonathan M.H. Gosling}
\author{Markus Rademacher}

\affiliation{Department of Physics \& Astronomy, University College London, London WC1E 6BT, UK}
\author{Jence T. Mulder}
\author{Arjan J. Houtepen}
\affiliation{Optoelectronic Materials Section, Faculty of Applied Sciences, Delft University of Technology, van der Maasweg 9, 2629 HZ Delft, The Netherlands}
\author{Marko Toroš}
\affiliation{School of Physics and Astronomy, University of Glasgow, Glasgow, G12 8QQ, UK}
\author{A.T.M. Anishur Rahman}
\affiliation{Department of Physics \& Astronomy, University College London, London WC1E 6BT, UK}
\author{Antonio Pontin}
\author{P. F. Barker}
\affiliation{Department of Physics \& Astronomy, University College London, London WC1E 6BT, UK}

\begin{abstract}
Fast detection and characterization of single nanoparticles such as viruses, airborne aerosols and colloidal particles are considered to be particularly important for medical applications, material science and atmospheric physics. In particular, non-intrusive optical characterization, which can be carried out in isolation from other particles, and without the deleterious effects of a substrate or solvent, is seen to be particularly important. Optical characterization via the scattering of light does not require complicated sample preparation and can in principle be carried out in-situ. We describe the characterization of single nanoparticle shape based on the measurement of their rotational and oscillatory motion when optically levitated within vacuum. Using colloidally grown yttrium lithium fluoride nanocrystals of different sizes, trapped in a single-beam optical tweezer, we demonstrate the utility of this method which is in good agreement with simulations of the dynamics. Size differences as small as a few nanometers could be resolved using this technique offering a new optical spectroscopic tool for non-contact characterization of single nanoparticles in the absence of a substrate.  
\end{abstract}
\maketitle
\section{Introduction}
The use of single-particle optical trapping has proven to be an important aid in nanoparticle characterization allowing for more time to fully characterize the light scattered from the nanoparticle and for detailed measurements of the nanoparticle's interaction with the surrounding environment providing information on the size and shape \cite{wu2017stable,kohler_tracking_2021,li2021fast}. Optical tweezer experiments have almost exclusively been undertaken in the environment of a liquid where the motion of the trapped particle is strongly damped \cite{Marago2010Brownian}. This masks much of the dynamics in the optical potential which is an important resource for detailed characterization which naturally emerges in the underdamped region of levitated optomechanics.

In the last decade, single nanoparticles have been levitated in gaseous environments or even in high vacuum \cite{Ballestro2021Levitodynamics}. The center-of-mass motion, as well as the rotational and librational motion, can be well controlled using optical or electrical fields. In these underdamped environments, optical cooling has been used to reduce motional temperatures to well below 1 K and recently single trapped nanoparticles have been placed in their motional quantum ground state \cite{Delic2020,Magrini2021,Tebbenjohanns2021}. The rotational dynamics of trapped non-spherical particles have been characterized in these types of traps and variations in lineshape and librational motion attributed to particle morphology have been measured \cite{hoang2016torsional}. The simultaneous cooling of all six motional degrees of freedom, which include the translational and rotational motion has been accomplished \cite{pontin2023simultaneous,kamba2023nanoscale}.

This exquisite optical control, coupled with an understanding of both the underdamped center-of-mass motion and the rotational dynamics, offers a new tool for the characterization of isolated single nanoparticles.  We call this spectroscopic characterization, levitodynamic spectroscopy.  Like rotational or vibrational spectroscopy of molecules, which provides information on molecular structure via its light-induced motion \cite{Kroto2003Molecular},  we show that the frequency-resolved dynamics of the motion within an optical trap can also be used for highly sensitive characterization of nanoparticle structure.  We illustrate the utilisation of this technique by computing spectra for a range of different shapes and demonstrate this technique on well-characterised samples of colloidally grown nanocrystals. Excellent agreement between the computed spectra and our experiments indicate how this technique could be used for fast and accurate characterization of single isolated nanostructures, such as viruses. With further development, this technique could provide rapid, high-throughput, measurement of single nanoparticle morphology\cite{pan2019collection,mitra2010nano,WANG2022105880,Zhu:11,Simic2022}. This is akin to that which is typically attained with electron microscopy or diffractive methods, both of which find wide application in pharma, electronics, energy storage, and materials discovery sectors.
\section{Levitodyanmic spectroscopy}
\begin{figure}
\includegraphics[width=\columnwidth]{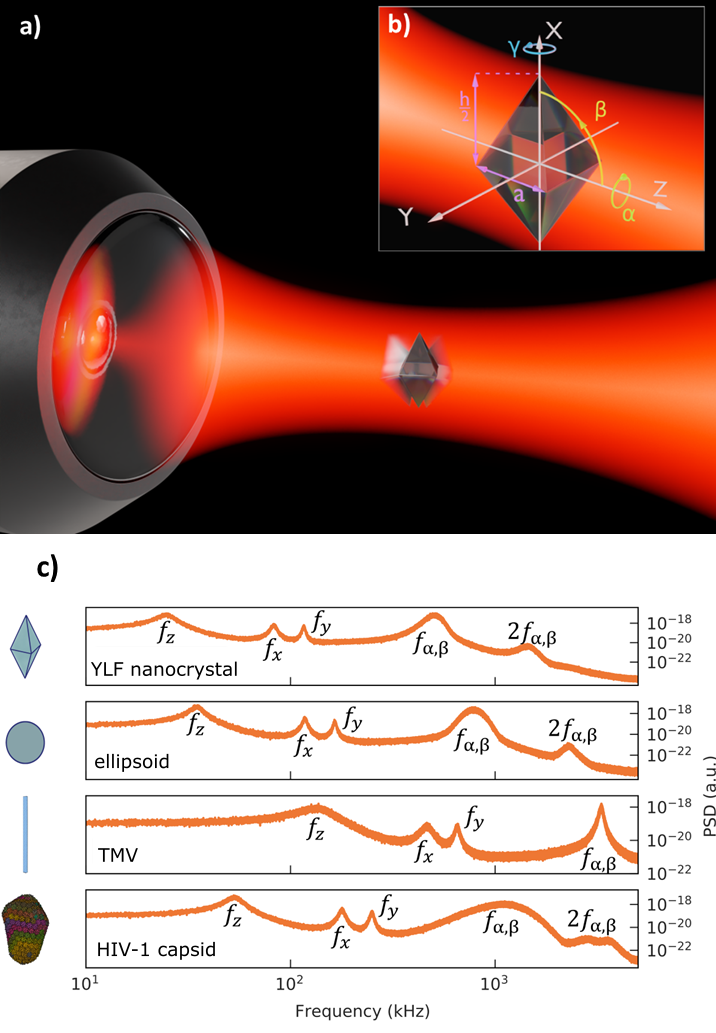}
\centering
\caption{\label{fig:Particle}a) Octahedral anisotropic particle levitated at low pressure undergoing translational and librational motion in a linearly polarized optical tweezer field. b) The undamped motion in the optical field undergoes translational ($x, y, z$) and librational motion ($\alpha, \beta, \gamma$) which can be measured from the scattered light.  Also shown are the dimensions of the crystal base and height. c) Simulations of the levitodynamic spectra derived from the motion of all six degrees of freedom which can be recorded on a single detector.  The calculations are shown for an octahedral yttrium lithium fluoride nanocrystal, silica ellipsoid, tobacco mosaic virus (TMV) and a hollow HIV capsid structure. The spectra illustrate the potential for using these spectral features of the underdamped motion as a tool of characterization of single nanoparticles.}
\end{figure}

Optically levitated anisotropic nanoparticles have three translational degrees of freedom $x$, $y$ and $z$ and three rotational degrees of freedom $\alpha$, $\beta$ and $\gamma$ which can be controlled via optical fields as shown in Fig. \ref{fig:Particle} a) $\&$ b) for an octahedral nanocrystal. A common method of optical levitation is to use an optical tweezer which is created by strongly focusing a Gaussian laser beam in a high numerical aperture lens. To illustrate the usefulness of this technique we describe the interaction of the optical field with the nanoparticle within the dipole approximation, where the nanoparticle size is significantly smaller than the wavelength of light used. A nanoparticle in an optical field is subject to both gradient and scattering optical forces which can act to strongly couple these degrees of freedom as has already been demonstrated in a range of experiments \cite{kuhn2017full,ahn2018optically,hoang2016torsional,Winstone2022Optical,arita2013laser,rahman2017laser}. When nonspherical nanoparticles are levitated within a tightly focused Gaussian optical field with linear polarization aligned along the $x$ axis of Fig. \ref{fig:Particle} b), the field creates a confining optical potential for both the center-of-mass motion as well as some of the rotational degrees of freedom and is dependent on the shape of the nanoparticle. Trapping leads to three distinct oscillation frequencies, which for example along the y-axis in the center of the optical potential for a symmetric top shape is given by $\omega_y=\sqrt{\frac{2 I_0 \chi_3}{ \rho \epsilon_0 c w_0^2}}$. Here the peak intensity of the Gaussian optical field is given by $I_0$, $\chi_3$ is the susceptibility along the longest axis, $\rho$ is the density of the nanoparticle, $w_0$ is the beam waist and $\epsilon_0$ the permittivity of free space. Similar relations are found for the other two axes and for different shapes \cite{hoang2016torsional}. Note that the oscillation frequency is proportional to square root of the susceptibility-to-density ratio $\chi_3/\rho$ along the longest axis of the nanoparticle which is optically aligned along the $x$-axis.  The oscillation frequency around the $\alpha$ rotational axis, where $\alpha\approx0$, is $\omega_{\alpha}=\sqrt{\frac{I_0 V (\chi_3-\chi_1)}{\epsilon_0 c J_1 }}$ which is proportional to the square root of the susceptibility difference to the moment of inertia $J_1$ ratio. Here, $V$ is the volume of the nanoparticle. A similar relation can be found for the $\beta$ rotational axis.  As the susceptibility is determined by both the nanoparticle composition and its dimensions, it offers the potential to utilise a measurement of these frequencies, and their variation, to determine nanoparticle shape \cite{rademacher2022measurement}. Rotational motion in general offers even more selectivity as it is particularly sensitive to the difference in susceptibility and the moment of inertia which is related to the difference in nanoparticle shape along its axes.
%\begin{figure}
%\centering
%\includegraphics[width=\columnwidth]{figs/Sim_placeholder.png}
%\begin{center} 
%\footnotesize \hspace{4 mm} 
%\end{center}
%\end{figure}

Whilst the interaction of the nanoparticle with the optical field provides information on its structure via the trap frequencies, so too does its interaction with the surrounding gas~\cite{cavalleri2010gas,epstein1924resistance}. In the weak damping limit, encountered when an anisotropic nanoparticle is optically trapped and aligned at low pressure, the damping in each direction is different, due to the difference in effective nanoparticle area in each direction of motion. This is experimentally resolvable when the surrounding gas pressure is in the millibar range. Thus, the librational and translation frequencies, as well as their frequency spread provides a spectral signature of the trapped nanoparticle. 

To illustrate the variation in the spectra for a range of species we solve the system of 12 coupled stochastic differential equations that describe the motion of an optically trapped anisotropic particle. This takes into account the three translational degrees of freedom $x$, $y$ and $z$ as well as the three rotational degrees of freedom $\alpha$, $\beta$ and $\gamma$ of the nanoparticle in the laboratory frame \cite{rashid2018precession, pontin2023simultaneous}. To describe the interaction of the optical field and the nanoparticle we use the dipole approximation where the particles are significantly smaller than the wavelength of light used to trap them. To predict the spectrum for a particular optical potential, and the environmental parameters in the lab, we consider the laser power, laser wavelength and the asymmetry of the optical potential created by a focused optical field. The anisotropic nanoparticle is described via its moments of inertia, damping coefficients and its dielectric susceptibilities along its principal axes. In addition, the nanoparticle's mass and volume as well as the gas pressure, the mass of the gas molecules and the gas temperature are also required. 

%the potential energy of this particle in a linearly polarized optical field can be approximated by
%\begin{equation}
%U(x,y,x,\alpha,\beta)=-\frac{1}{4}I_0\frac{V_s}{c(1+\frac{z^2}{zr^2})}[(\chi_1+x2+(\chi1-\chi3)\cos{2 \beta})\cos^2{\alpha}+\chi_1 %\sin^2{\alpha}] \exp{-2\frac{x^2}{w_x(z)^2}-2frac{y^2}{w_y(z)^2}  
%\end{equation}
The translational and rotational motion of a range of organic and inorganic nanoparticles are simulated, ranging from octahedral yttrium lithium fluoride (YLF) nanocrystals and silica ellipsoids to the tobacco mosaic virus (TMV) and HIV capsids, all of which are in the 100 nm size range but with different morphologies, see Fig. \ref{fig:Particle} c). Further details on the particle structure and optical fields used in the calculation are given in the methods section below.  We calculate the evolution of all six degrees of freedom in a linearly polarized field as a function of time for a duration that is much longer than the largest characteristic damping time at the pressure of interest. We subsequently determine the power spectral density (PSD) of each degree of freedom to generate the levitodynamical spectra. As all of these degrees of freedom can be measured on a single detector, which is typically used for rotation measurement, we add all three of these features together such that both the rotational and translational features have approximately the same amplitude as observed in experiments. The relative spectral magnitude of these features in any experiment will depend on the detailed detection process and efficiency. However, these calculated spectra serve to illustrate the significant variation in the spectral location and width (lineshape) of the translational and librational features which could be used to identify, characterise and differentiate between different single levitated nanoparticles. Labelled on Fig. \ref{fig:Particle} c) are the peaks which relate to motion in the $x$, $y$ and $z$ direction, as well as the $\alpha$ and $\beta$ rotational motion for these particles. For these particles, which can be considered symmetric top rotors, the motion along $\gamma$ is not directly coupled to the optical field but diffuses freely via collisions with the background gas. It is however coupled to the other two rotational degrees of freedom and leads to a diffusion in their librational frequencies which acts to broaden these spectral features \cite{bang2020five}.  Note that at the pressures and intensities used in our experiments, and for most of the optical tweezer experiments considered here, the spectral widths of the translational features are determined by the gas damping and librational alignment, whilst the broadening of the rotational peaks is determined by the degree of alignment of the particles in the optical fields and the non-linear coupling between rotational degrees of freedom.

\section{Spectroscopy of Octahedral Nanocrystals}
\begin{figure}
\includegraphics[width=1\columnwidth]{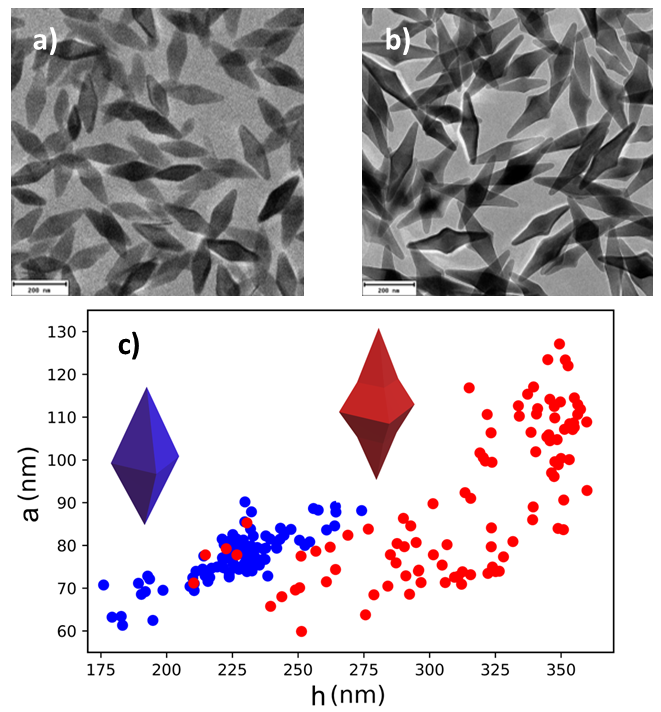}
\centering
\caption{\label{fig:TEM} a) TEM image of sample 1, long side length of \textbf{$(227 \pm 18)$ nm}, short side length of \textbf{$(77 \pm 6)$ nm}. b) TEM image of sample 2, long side length of \textbf{$(314 \pm 38)$ nm}, short side length of \textbf{$(91 \pm 18)$ nm}. c) Measurements of height $h$, and base width $a$ for both sample 1 shown in blue and sample 2 shown in red for one hundred measurements taken from TEM images showing the distribution of sizes in these colloidally grown octahedral nanocrystals.}
\end{figure}

To demonstrate this spectroscopic method, and compare with simulations, we measure the PSD acquired for a well-defined nanoparticle structure.  For this we use colloidally grown octahedral YLF nanocrystals, whose synthesis has been described elsewhere \cite{mulder2023nucleation,mulder2023understanding}. We determined the dimensions of two different samples of these nanocrystals by imaging with a transmission electron microscope (TEM). TEM images of the nanocrystal ensembles are shown in Fig. \ref{fig:TEM} a) \& b).  From a hundred imaged nanocrystals we determine the mean height, $h$, width, $a$, as well as the standard deviation of the measured distribution. For sample 1, $h$ = 227$\pm$18 nm, $a$ = 77$\pm$6 nm and for sample 2  $h$ = 314$\pm$38 nm , $a$ = 91$\pm$18 nm. The full distribution of $a$ and $h$ is given in Fig. \ref{fig:TEM} c) and shows that sample two generally has higher aspect ratios $h/a$. 

\begin{figure}
\includegraphics[width=1\columnwidth]{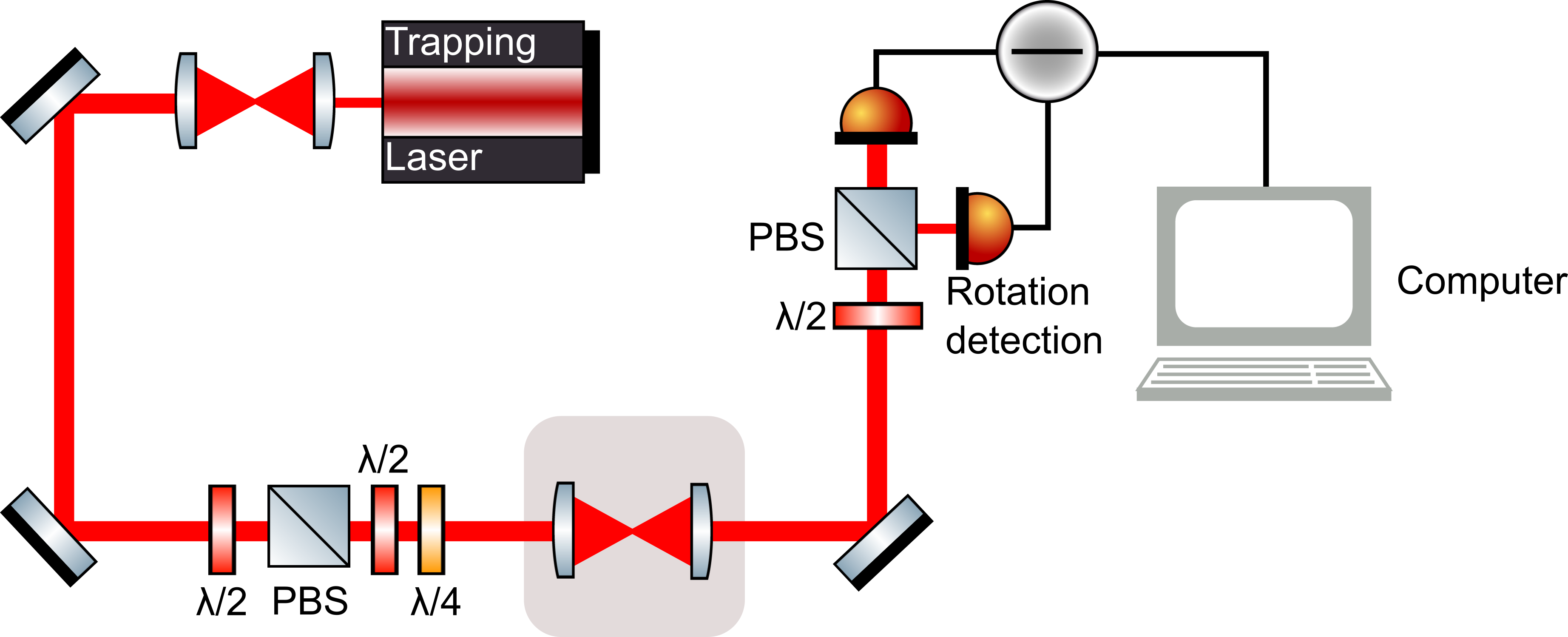}
\centering
\caption{\label{setup} Experimental layout used throughout this study. The setup consists of an optical tweezer formed by tightly focusing a 1030 nm laser beam with an aspheric lens, NA=0.77. The scattered light is collected by a collimation lens and directed towards three balanced photodetectors which are used to measure the translational and librational motion.}
\end{figure}

We record the levitodynamic spectra for the two sizes of the colloidally grown YLF nanocrystals using an optical tweezer in vacuum at a pressure of 5 mbar.  The optical tweezer was formed by tightly focusing a linearly polarized 1030 nm laser beam of power 200 mW with a NA=0.77 aspheric lens. The librational motion of the levitated nanocrystal was detected by observing changes in polarization in the forward transmitted beam which includes light scattered by the nanoparticle \cite{Nieminen2000}. This is particularly sensitive to the motion in the $\alpha$ degree of freedom but also enables us to detect the translational motion. Note that we do not observe the lower frequency motion in the $z$ direction as we use a highpass filter to suppress low frequency noise.  The small change in polarization needed to observe the rotational motion is detected using a polarizing beam splitter which converts the polarization change into intensity modulation \cite{rahman2017laser,hoang2016torsional}. A schematic of the experimental layout can be found in Fig. \ref{setup}.


\begin{figure}
\includegraphics[width=\columnwidth]{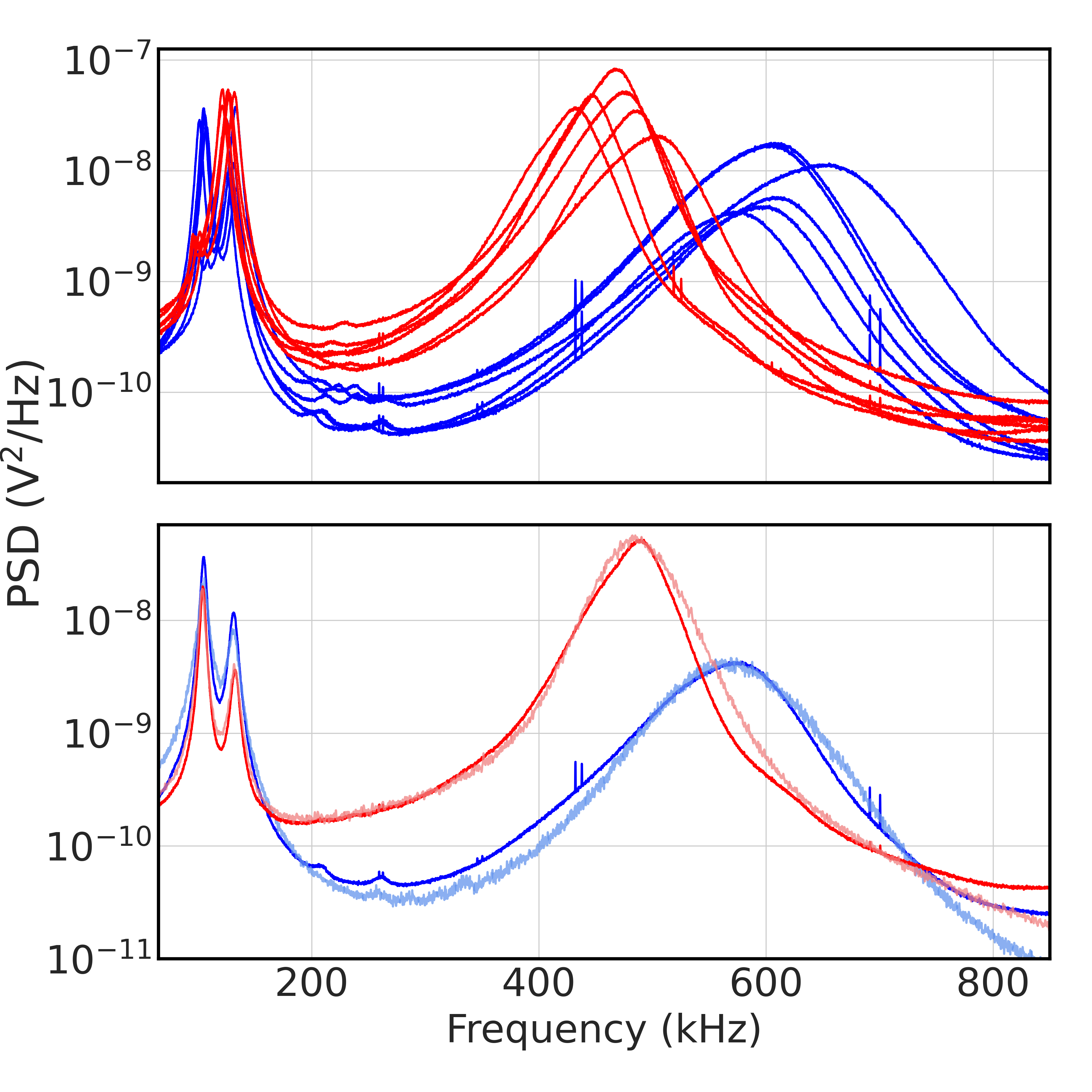}
\centering
\caption{\label{fig:PSD} The upper plot shows the levitodynamic spectra for six levitated nanocrystals taken from samples 1 (blue) and 2 (red). The frequencies for translational motion in the $x$ and $y$ directions are at approximately 95 kHz and 120 kHz respectively. The frequencies of the librational motion are in the region of 430-510 kHz for sample 2 and 570-650 kHz for sample 1. While there is little difference in the translational frequency, there is a clear difference between librational frequency peaks centering at approximately 500 kHz and 600 kHz.  The thin spikes observed experimentally are due to electronic noise in the detection scheme. The lower plot shows a fit (light thicker line) of the levitodynamic spectra obtained from the simulation of the evolution of the nanocrystal’s motion with respect to time to the experimental data (dark thin line).}
\end{figure}

% \begin{figure}
% \includegraphics[width=1\columnwidth]{figs/Rotation detection setup.pdf}
% \centering
% \caption{\label{fig:layout} Experimental layout used throughout this study. It consists of an optical tweezer formed by tightly focusing a  1030 nm laser beam with an aspheric lens, NA=0.77. The scattered light is collected by a collimation lens and directed towards three balanced photodetectors which are used to measure the translational and librational motion.}
% \end{figure} \\
Fig. \ref{fig:PSD} shows PSDs for six crystals for each of sample 1 and sample 2. These are obtained from the Fourier transforms of the time series of 32 data sets using a temporal window of 500 ms sampled at 2 MHz. All spectra show three well-defined peaks, the two lower frequency, narrower peaks are due to translational motion in the direction transverse to the propagation of the beam, denoted $f_{x}$ and $f_{y}$.  The broader peak at high frequencies corresponds to the librational frequency, denoted $f_{\alpha}$, which can also show information on $f_{\beta}$, broadened by the diffusion of the librational frequency around a central value, as has been observed elsewhere \cite{hoang2016torsional,kuhn2017full,zielinska2023full,kamba2023nanoscale,pontin2023simultaneous,vanderLaan2020Optically,rashid2018precession}. The very thin spikes are due to electronic noise in the detection. There is a large difference in the librational frequencies between the two samples, but there is also a difference in frequency within the samples. Additionally, the width of these features varies between each sample. This illustrates the potential of levitodynamic spectroscopy and the use of librational motion to differentiate between nanoparticle size even for similar morphology.

 % \begin{figure}[h!]
 % \begin{minipage}[b]{1\columnwidth}
 % \centering
 % \includegraphics[width=1\columnwidth]{figs/Sim_vs_experiment_sample_1_2_particles.pdf}
 % \begin{center} 
 % \footnotesize \hspace{4 mm} 
 % \end{center}
 % \end{minipage}
 % \hspace{0.9cm}
 % \begin{minipage}[b]{\columnwidth}
 % \centering
 % \includegraphics[width=\columnwidth]{figs/ComparisonUnadjust.pdf}
 % \begin{center} 
 % \footnotesize \hspace{4 mm} 
 % \end{center}
 % \end{minipage}
 % \hspace{0.9cm}
 % \begin{minipage}[b]{\columnwidth}
 % \centering
 % \includegraphics[width=\columnwidth]{figs/ComparisonSim.pdf}
 % \begin{center} 
 % \footnotesize \hspace{4 mm} 
 % \end{center}
 % \end{minipage}
 % \hspace{0.9cm}
 % \caption{\label{fig:figure6}(a) The power spectral density at 5 mbar for a linearly polarized tweezer beam for sample 1 and 2. The frequencies for translational in the $z$, $x$, $\textup{and}$ $y$ are at approximately 25 kHz, 95 kHz and 120 kHz respectively. The frequencies of the librational motion are in the region of 420-450 kHz for sample 2 and 530-580 kHz for sample 1. The smaller peaks at 800-850 kHz are the second harmonics of the rotational motion. (b) A histogram showing the distribution of the librational frequencies for approximately twenty nanocrystals of sample 1 and 2 levitated in the trap.(c) A histogram showing the theoretical distribution of the librational frequencies for approximately one hundred nanocrystals of sample 1 and 2 from the SEM measurements.}
 % \end{figure} 

We now compare these experimental levitodynamic spectra with simulated spectra derived from the stochastic simulations of the translational and librational motion of the nanocrystals in the trap. We calculate the susceptibility tensor, moment of inertia, mass and volume as a function of the particles height $h$ and the base $a$ assuming a simple octahedral structure for sample 1 and octahedral structure with a central bulge for sample 2 as observed in the TEM images. This is combined with a measurement of laser power and gas pressure. The simulated spectra are shown in the lower plot of Fig. \ref{fig:PSD} overlapped with experimental spectra from each sample. The light thicker line represents the simulated fit to the levitodynamic spectra of the dark thin lined experimental data. 

As there is some variation between nanocrystal size in each sample, the calculated spectra for each was derived from a manual fitting procedure to estimate the nanocrystal dimensions. This fitting procedure is informed by our understanding of the experimental uncertainties in the focused laser beam, but also from the bounds of nanocrystal dimensions set by TEM images of over a hundred nanocrystals in each sample. We initially match the position of the translational frequencies by varying the focused beam dimensions including the beam asymmetry and its estimated power based on the input power. As the librational spectral features are more sensitive to the nanocrystal's dimensions than the translational features, we first scale the $a$ and $h$ dimensions of the octahedral structure equally to match the librational peaks of the $\alpha$ motion with the experimental data. Subsequently, we scale the $a$ dimension independently of $h$ to match the line width ratios measured in the experiment. Finally, we adjust the relative amplitude of the PSD from each degree of freedom to take into account the variation in detection efficiency for each degree of freedom. We iteratively repeat this process to reach the best fit. 
% We note that this fitting procedure maintains the same optical field properties for both trapped samples.  

The good fit between the experiment and simulation across all particles in both samples illustrates its potential for single nanoparticle characterization. For sample 1, the goodness of fit $\chi^2 = 3 \times 10^{-6}$ and for sample 2, $\chi^2 = 5 \times 10^{-5}$.
%For sample 1, $\chi^2 = 2 \times 10^{-5}$ and for sample 2, $\chi^2 = 5 \times 10^{-5}$. This considers the whole PSD spectra. 

Here the line widths of the translational peaks provide information on the aspect ratio ($h/a$) of the octahedron, while the position and the width of the librational spectral feature are particularly sensitive to the absolute value of these dimensions.  Note that the simulations produce a broader librational peak which could be attributed to the breakdown in the dipole approximation. Based on the good fit of our simulations we can estimate how well we can differentiate between size. As we can straightforwardly determine the difference between librational peak shifts of 1 kHz this corresponds to a change in either the $a$ or $h$ dimension of both structures by approximately 1 nm. 

%For these simulations we obtain the value of the gas damping of the translational and librational motion using a numerical procedure derived from refx[]. These values and the mean size of the nanocrystals derived from SEM images serve as a starting point for a manual fitting procedure.
%When comparing the spectra from sample 1 and 2 we can see that there is a greater separation in the frequency of the librational peaks (approximately 100 kHz) when compared to the translational motion (approximately 3 kHz).%

%While we have shown only a few of the recorded spectra 
%we show the distribution of peak frequency of the librational motion recorded for twenty two crystals in %each sample. These measurements are shown in Fig.\ref{fig:figure6} b. \\

Our measurements show that the average librational frequency distribution varies significantly between the two nanocrystals samples due to the distribution in crystal size. Here we consider the spectra of 13 particles from sample 1 and 25 from sample 2. For sample 1, the experimental distribution of the librational frequencies has an average frequency of 600 kHz with a standard deviation of 3\%.  This spread is narrower than sample 2 with an average frequency of 500 kHz and standard deviation of 10\%. The larger variation in sample 2 corresponds well to the wider variation observed from the TEM imaging. Here the variance in height was 12 \% while the width was 20 \%. This compares with approximately 8\% for both dimensions of sample 1.

\begin{figure}[h!]
\begin{minipage}[b]{1\columnwidth}
\centering
\includegraphics[width=1\columnwidth]{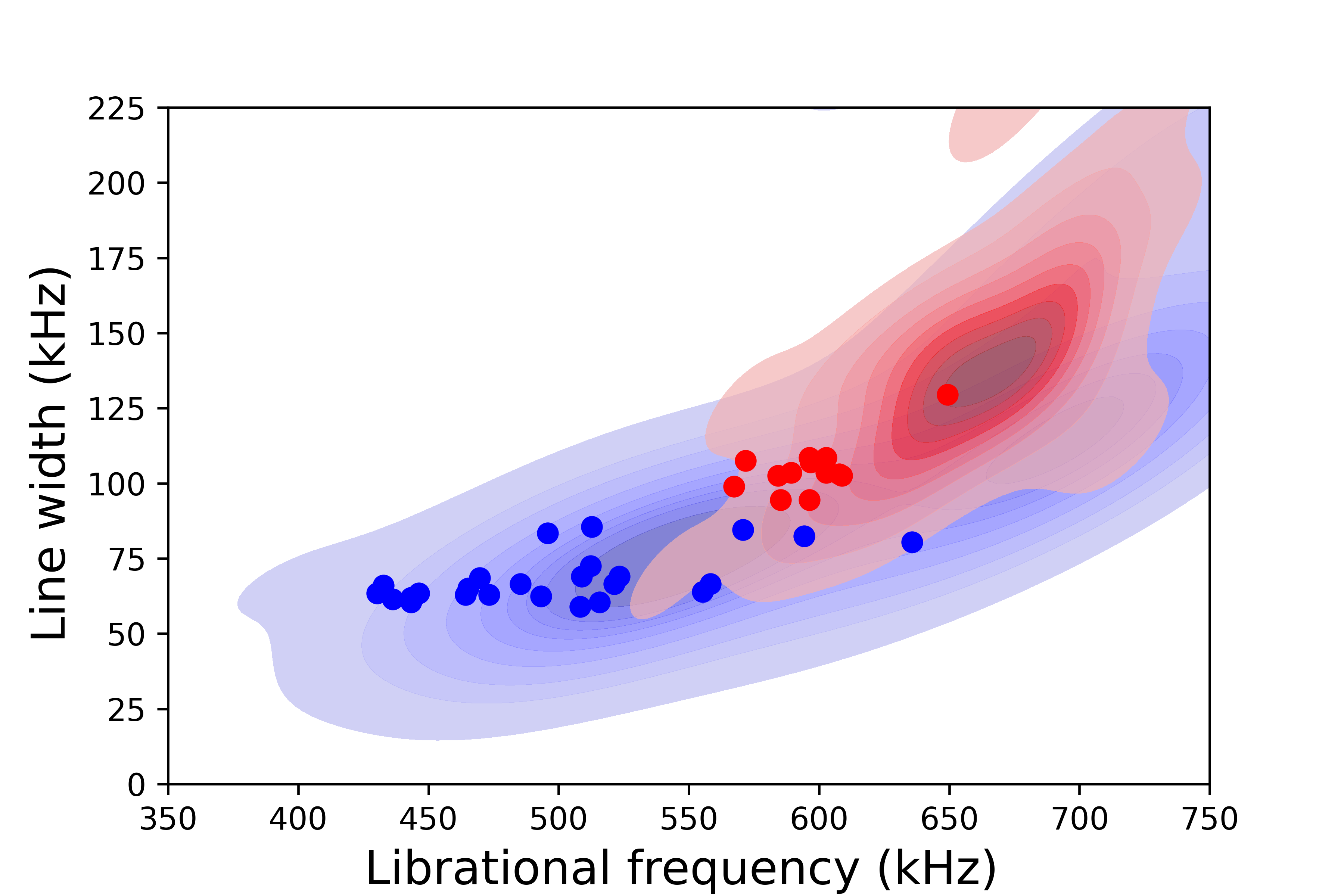}
\begin{center} 
\footnotesize \hspace{4 mm} 
\end{center}
\end{minipage}
\caption{\label{fig:ratio}Librational line width (full-width-half maximum) versus librational frequency for samples 1 (blue) and 2 (red). Experimental data points are shown as circles while the results from 100 simulations derived from the TEM data measurements of crystal dimensions $h$ and $a$ are shown as a two-dimensional density plot where blue contours represent simulations for sample 1 and red for sample 2. The experimental librational frequencies being lower than the simulations are likely due to smaller nanocrystals being unable to be trapped.}
\end{figure}

We now compare the experimental librational and translational spectra from each of the two colloidally grown samples with simulations based on the TEM measurements of $h$ and $a$. This simulation takes into account the deviations from the octahedral shape by the central bulge observed in sample 2. Here we approximate sample 2 as two composite intersecting octahedral structures where one has height $h/1.3$ and base $a$, and another with height $h$ and base $a/0.8$. To make this comparison, we plot the full-width-half-maximum, around the peak of the broad librational feature, as a function of this peak value. This is shown in Fig. \ref{fig:ratio}, where the circles are the experimental data derived from the spectra of 13 particles from sample 1 (blue) and 25 from sample 2 (red). It is clear from this plot that the two samples are well separated which allows differentiation from each other. 

To compare with the simulations of librational line width versus librational peak frequency for each of the 100 particles in each sample, we instead provide a 2-D density plot derived from the kernel density estimation (KDE). It is clear that the data recorded in the experiment overlaps well with the lower librational frequencies and line widths predicted for each sample set in the lower left of the figure but not at higher frequencies and line widths. The agreement between experiment and the simulations in the lower left part of the plot can only be achieved by taking into account the changes in the moment of inertia and susceptibility due to the central bulge of sample 2. However, it is clear that there are no recorded spectra with higher librational frequencies and line widths that match the simulations in the upper right of this figure. These correspond to particles that have smaller aspect ratios $h/a$, but more importantly have smaller volumes and polarizabilities which lead to lower optical well depths when compared to other particles in the distribution. These particles are less likely to be initially captured by the optical trap and to stay trapped even if captured and most likely accounts for the absence of particles at the higher librational frequencies and line widths.   
%\
%(same trap with different sizes good.)\\

\begin{figure}[h!]
\begin{minipage}[b]{1\columnwidth}
\centering
\includegraphics[width=1\columnwidth]{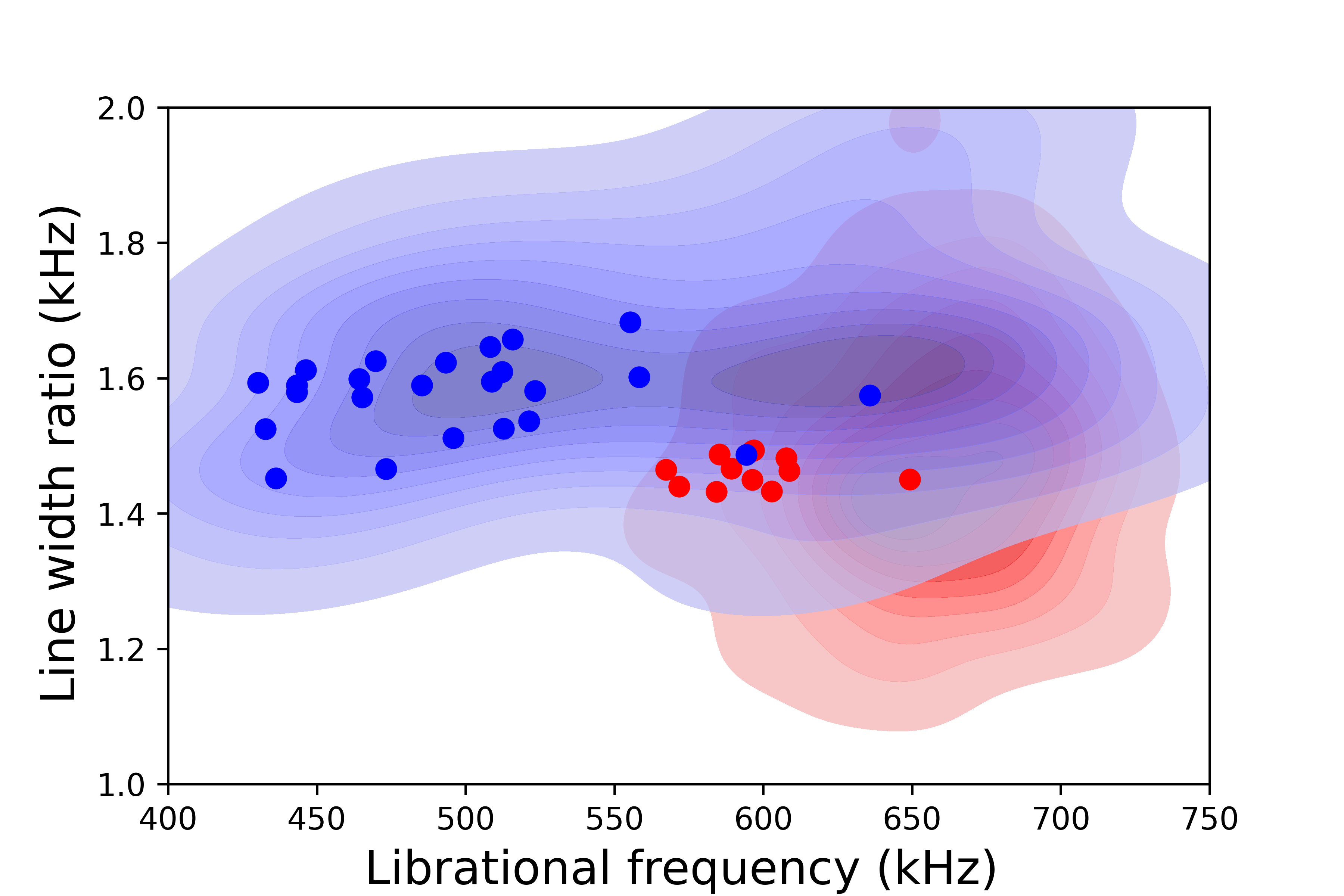}
\begin{center} 
\footnotesize \hspace{4 mm} 
\end{center}
\end{minipage}
\caption{\label{fig:Linewidth}The distribution of the translational linewidth ratio $\gamma_{y}/\gamma_{x}$ as a functions of librational frequencies for sample 1 (blue) and 2 (red). Experimental data points are shown as circles while the results from 100 simulations derived from TEM data are shown as a two dimensional density plot where blue contours represent simulations for sample 1 and red for sample 2. The experimental librational frequencies being lower than the simulations are likely due to smaller nanocrystals being unable to be trapped.}
\end{figure}

The ratio of the translational motion's line width $\gamma_{y}/\gamma_{x}$ is strongly dependent on the particle anisotropy ratio $h/a$. The translational line widths are determined by a fit to the PSD's of this degree of freedom. When aligned by the field the particles experience different gas damping in the $x$ and $y$ directions \cite{cavalleri2010gas,hoang2016torsional}. For nanocrystals that are longer and thinner in shape, the difference between the gas damping becomes significant. The variation of the spectra of the two samples is summarised in  Fig. \ref{fig:Linewidth} which plots the translational line width ratio versus the central frequency of the librational motion.  While there is some overlap of the line width ratios, when the librational frequencies are taken into account, the two samples are clearly separated on the plot allowing us to clearly differentiate between the two types of samples.  The average line width ratios are 1.46$\pm$0.02 and 1.58$\pm$0.06 for sample 1 and sample 2 respectively. As expected, sample 1 which has the lower aspect ratio has the lower line width ratio, which is also confirmed with our simulations based on the dimensions measured from the TEM images. This give values of 1.59$\pm$0.05 and 1.76$\pm$0.08 for sample 1 and 2 respectively. Also shown are KDE contour plots derived from the spectral simulations of 100 particles in each sample. Again the higher librational frequencies that occur in the simulations are absent in the data probably due to the inability to trap these smaller particles for the duration required to measure the spectra of the trapped particles.   
  
%There is a clear separation between the two nanocrystal types with clustering in two distinct %positions. It also illustrates the power of using two discriminating methods for nanocrystal %characterization. In the small dataset taken, there are overlaps of nanocrystal linewidth %ratios for the two sample types. Whilst the aspect ratios for the two nanocrystals are similar, %as they come from two different sample types, their overall size is likely to be different. %This is highlighted by the large separation between the nanocrystal librational frequencies. %Being able to discriminate between these two nanocrystal types would be very difficult using %one technique alone.

\section{Conclusions}
We developed a technique called levitodynamic spectroscopy where the spectral peaks and line widths associated with both the translational and librational degrees of freedom of a single trapped nanoparticle can be used as a particle-specific high-resolution fingerprint. Using nanocrystals with a well-defined morphology we have shown that the spectral features are not just dependent on the shape, but also on the density and mass distribution, as well as the optical susceptibility of the particles. 

The potential of this approach to particle characterization is illustrated by a comparison between our levitodynamic spectra and simulations. The simulations were based on the measured properties of two ensembles of colloidally grown bipyramidal yttrium lithium fluoride nanocrystals which have a well-defined shape, susceptibility and density. While we have studied small particles which are relatively well described within the Rayleigh approximation, our models could be modified to operate in the Mie regime \cite{PhysRevA.108.033714},  which would allow us to explore levitodynamic spectroscopy for micron and larger sized particles.  Like conventional molecular rotational spectroscopy, levitodynamic spectroscopy reveals information on nanoparticle properties via the rotational dynamics.  Levitodynamic spectroscopy additionally provides information via the oscillatory nature of the translational trapped motion within the optical potential. This is analogous to molecular vibrational Raman spectroscopy which probes a molecules interatomic potentials. For the well-defined particles studied here, the excellent comparison between our experiments and simulations suggest that differences as little as a few nanometers along any direction could be differentiated using this technique. 

While further work needs to be undertaken to establish the general utility of this method for a wide range of nanoparticle morphologies, our simulations of levitodynamic spectra of virus particles and inorganic nanocrystals suggest that a wide range of particles could be identified using this method. Since even similar shaped particles are likely to have different density, mass distribution and optical properties, there is significant promise for obtaining unique fingerprints of a large range of particles using levitodynamic spectroscopy. The ability to model the spectra for different morphologies, as demonstrated here, will be key to identification via this technique. Like many complex spectroscopic measurements, spectral features are often simply compared to a database for identification and is amenable to the application of machine learning with input from both measurements and modelling.  This approach would however require a standard optical trapping potential and a well-defined gas pressure for operation. This spectroscopic approach is very general as a wide range of particles can be trapped at gas pressures in the few millibar regime. At this pressure the gas damping is low enough so that spectral features are well resolved, while laser induced heating can be controlled by the collisional cooling from the surrounding gas. This prevents significant heating and loss of nanoparticles from the trap at lower pressures, allowing the trapping of more strongly absorbing particles (gold nanorods) as well as organic particles such as viruses (granuloviruses and tobacco mosaic viruses).  

Finally, while we have explored levitodynamic spectroscopy within linearly polarized optical fields, elliptical or even circularly polarized fields could also be used to reveal similar or even additional structural information \cite{hu2023structured,ahn2020ultrasensitive,reimann2018ghz,mazilu2016orbital}. Similarly other optical traps could be used to extend the characterization to smaller and larger particles by utilising standing wave or counter-propagating beam traps.  Lastly, levitodynamic spectroscopy when combined with angularly resolved scattering in the same setup could be used to provide even greater differentiation for similarly shaped nanoparticles \cite{rademacher2022measurement}.  
\section*{Acknowledgement}
J.G., M.R., A.R., A.P., and P.F.B. acknowledge funding from the
EPSRC via Grant Nos. EP/S000267/1 and EP/W029626/1. 
J.G., M.R., A.R., A.P., J.T.M., A.J.H, and P.F.B. acknowledge funding from the
H2020-EU.1.2.1 TEQ Project through Grant Agreement ID 766900.
M.T. acknowledges funding by the Leverhulme Trust (No. RPG-
2020-197).

%YLF nanocrystals are produced by a colloidal synthesis method where complexes of Li-oleate, and Y-oleate react with fluoride by cracking of trifluoroacetate at temperatures above 250 $^{\circ}$C. This results in the formation of YLF nuclei, which grow into bipyramidal shaped nanocrystals. This synthesis yields monodisperse, bipyramidal, nanocrystals that are colloidally stable with a much higher reproducibility. The oleate surfactants can be removed and replaced by BF4-. This prevents unwanted light absorption by the organic ligands and renders the colloidal particles stable in methanol.

%Open questions:
%\begin{enumerate}
%    \item How to best present the size distribution in the librational motion with the simulation? Stacking up PSD traces gets convoluted quite easily. 
%    \item What do we think about using Figure 2c from~\cite{hoang2016torsional} as layout template for figures depicting x-y linewidth ratios and scattering intensities?
%\end{enumerate}
%%

\bibliographystyle{apsrev4-2} % Tell bibtex which bibliography style to use
%\bibliography{references.bib}

%

%%%%%%%%%% Merge with supplemental materials %%%%%%%%%%
\clearpage
\newpage

\begin{center}
\textbf{\large Methods}
\end{center}

\section*{Nanoparticle susceptibility}
We numerically calculate the three directional susceptibilities $\chi_{zz}$, $\chi_{yy}$ and $\chi_{xx}$ of non-ellipsoidal symmetric top objects by determining the scattered intensity in the Rayleigh regime \cite{miles_laser_2001} using the finite-difference time-domain (FDTD) method~\cite{yee_numerical_1966,taflove_application_1980}. The nanoparticle shape and refractive index including any birefringence is included in the FDTD calculation. The scattered intensity in the far field provides access to $\chi_{zz}$ by rotating the object in reference to the incoming light polarization along the z-axis~\cite{miles_laser_2001}. From $\chi_{zz}$ we then calculate $\chi_{yy}$ via the symmetric properties of the particles. Calculations performed in this way are in good agreement with reference~\cite{ahn_optically_2018}. 
\begin{figure}[h!]
\begin{minipage}[b]{1\columnwidth}
\centering
\includegraphics[width=1\columnwidth]{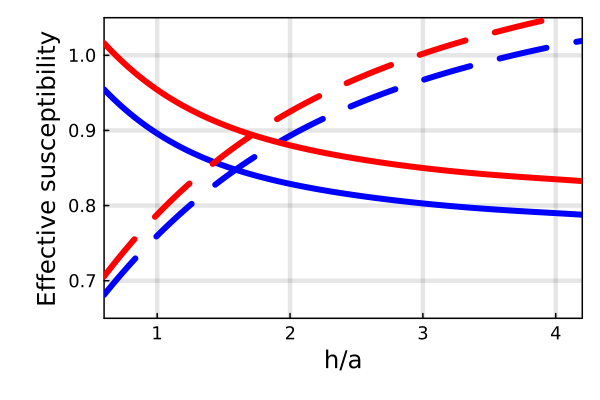}
\begin{center} 
\footnotesize \hspace{4 mm} 
\end{center}
\end{minipage}
\caption{\label{fig:Sratio}The susceptibility of sample 1 modelled as a octahedron of height $h$ and base length $a$ of sample 1 (blue) and and octahedron with a central bulge represented by the intersection of two octahedrons height $h$ and  sample 2. Each has an average refractive index of 1.45. }
\end{figure} 
For the octahedral YLF structures and the ellipsoid we use an average refractive index of 1.45 and for the biological structures we use a value of 1.47 \cite{Amin2019} (HIV capsid, tobacco mosaic virus).

\section*{Gas damping}
To compute the gas-damping coefficients of the objects for our numerical simulations, a finite element method was used. This approach is based on reference~\cite{cavalleri2010gas} which allows us to gain insight into the damping behaviour of nano-scale objects when analytical computation is not feasible. 

First, we cover the object in a mesh consisting of interconnected triangular elements. We compute the midpoint $\Vec{\mathfrak{v}}_\otimes$ of every mesh element and the orthogonal vector~$\Vec{\mathfrak{v}}_\perp$ for all the triangular mesh elements to account for the perpendicular force contribution to the object. This allows us to calculate the area $\mathcal{A}$ for each triangular polygon in the mesh. We then select a vector $\Vec{\mathfrak{v}}_\odot$  to represent the axis along which the force noise $S_F$ is computed. We use the unit vector along the z-axis $\Vec{e}_z$ in this example. We calculate the weighing factors $\zeta_\parallel = (1 - \Vec{\mathfrak{v}}_\odot \cdot \Vec{\mathfrak{v}}_\perp)^\frac{1}{2}$ and $\zeta_\perp = \Vec{\mathfrak{v}}_\odot\cdot\Vec{\mathfrak{v}}_\perp$. 
After accumulating these factors over the entire surface of the mesh, we obtain an expression for the total force noise $S_F=\sum_i \mathcal{A}_i (\zeta_{i,\perp}^2 S_\perp + \zeta_{i,\parallel}^2 S_{\parallel_{1,2}})$. We arrive at a total force noise and a translational damping coefficient $\beta_{\text{tr}}$ using $S_F = 4 k_B T \beta_{\text{tr}}$, which shows less than a 0.3\% discrepancy with the analytically computed values, validating the utility of our numerical approach.

Next, we calculate the torque noise $S_N$ for a chosen axis of rotation $\Vec{\mathfrak{v}}_\odot$  by forming an orthonormal basis for every midpoint $\Vec{\mathfrak{v}}_\otimes$ of each triangular mesh element. We compute the vectors, $\Vec{\mathfrak{v}}_{\parallel_1}$ and $\Vec{\mathfrak{v}}_{\parallel_2}$, parallel to the surface element using linear algebra. 

Afterward, we compute the weighing factors $\xi_\perp = \Vec{\mathfrak{v}}_\odot \cdot \left (\Vec{\mathfrak{v}}_\otimes \times \Vec{\mathfrak{v}}_\perp \right )$, $\xi_{\parallel_1}= \Vec{\mathfrak{v}}_\odot \cdot \left(\Vec{\mathfrak{v}}_\otimes \times \Vec{\mathfrak{v}}_{\parallel_1} \right)$, and $\xi_{\parallel_2}= \Vec{\mathfrak{v}}_\odot \cdot \left(\Vec{\mathfrak{v}}_\otimes \times \Vec{\mathfrak{v}}_{\parallel_2} \right)$ for every triangular mesh element and determine the total torque noise $S_N$ acting along a specific rotation axis $\Vec{\mathfrak{v}}_\odot$ by $S_N = \sum_i \mathcal{A}_i \left (\xi_{i,\perp}^2 S_\perp + \xi_{i,{\parallel_1}}^2 S_{\parallel_{1}} + \xi_{i,{\parallel_2}}^2 S_{\parallel_{2}} \right )$. We arrive at the rotational damping coefficient $\beta_{\text{rot}}$ using $S_N = 4 k_B T \beta_{\text{rot}}$.

The discrepancy between the numerically and analytically computed rotational damping coefficients is less than 0.8\% when compared with the analytical solution for a sphere.

\section*{Moment of inertia}
As an input to our simulations, we calculate the moment of inertia numerically for each particle around the center of mass. This provides three moments of inertia $J_1$, $J_2$ and $J_3$ around the body fixed axis. The density of the octahedral YLF crystals is taken as 3965 kg/m$^3$, the silica ellipsoid at 1850 kg/m$^3$ and the biological nanoparticles (HIV capsids and TMV) at 1325 kg/m$^3$.
\section*{Coupled equations of motion}
In the Hamiltonian formalism, we find 12 coupled equations of motion. The generalized position $q_i$ of the nanoparticle are given by three spatial positions ($x,y,z$) and three angular positions ($\alpha,\beta,\gamma$). For each position, a corresponding conjugate momenta are defined.  Considering standard Hamiltonian equations of motion, $\Ddot{q_i}=\partial_{p_i} H$ and $\Ddot{p_i}=-\partial_{q_i} H$, and a Hamiltonian of the form $H=T+V$, where $T$ and $V$ are the kinetic and potential energy, the twelve coupled stochastic equations of motions can be obtained:
\begin{equation}\label{eq:3.14}
    \textup{d}\textbf{r} =\partial_{p} H_{\textup{free}}  \textup{d}\textup{t},
\end{equation}
\begin{equation}\label{eq:32.15}
    \textup{d}\textbf{p} = - \partial_{r} H_{\textup{opto}} \textup{d}\textup{t}   + \textup{d}\textbf{p}^{(dc)}   + \textup{d}\textbf{p}^{(sc)},
\end{equation}
\begin{equation}\label{eq:3.16}
    \textup{d}\phi =\partial_{\pi} (H_{\textup{free}} + H_{\textup{opto}})  \textup{d}\textup{t},
\end{equation}
\begin{equation}\label{eq:3.17}
    \textup{d}\pi = - \partial_{\phi} (H_{\textup{free}} +H_{\textup{opto}}) \textup{d}\textup{t}  + \textup{d}\pi^{(dc)}  + \textup{d}\pi^{(sc)}
\end{equation}
where $\textbf{r}=(x,y,z)^\top$ is the spatial center-of-mass position, $\textbf{p}=(p_x,p_y,p_z)^\top$ is the center-of-mass conjugate momentum, $\phi=(\alpha,\beta,\gamma)^\top$ is vector of the Euler angles in the $z-y'-z''$ frame as described in Fig. \ref{fig:Particle} b) \cite{rashid2018precession} and $\pi=(\pi_\alpha,\pi_\beta,\pi_\gamma)^\top$ is the angular conjugate momentum. $dc$ denotes the non-conservative gas collision terms and $sc$ is the stochastic gas collision. $H_{\textup{free}}$ is the free Hamiltonian and $H_{\textup{opto}}$ is the Hamiltonian of the deterministic optomechanical terms.

To transform from the nanoparticle's reference frame ($x''$,$y''$,$z''$) to the laboratory reference frame ($x,y,z$), three rotations are applied, first around $z''$ (angle $\gamma$) then around $y'$ (angle $\beta$) and then finally around the $z$ axis (angle $\alpha$).

\clearpage

\end{document}